\PassOptionsToPackage{table,xcdraw,svgnames}{xcolor}
\pdfoutput=1
\documentclass[journal]{vgtc}                     % final (journal style)
\onlineid{0}

\vgtccategory{Research}

\vgtcpapertype{please specify}

\title{3D Gaussian Particle Approximation of VDB Datasets: A Study for Scientific Visualization}

\author{%
  \authororcid{Isha Sharma}{0000-0001-7978-3906},
  \authororcid{Dieter Schmalstieg}{0000-0003-2813-2235}
}

\authorfooter{
  \item
  	Isha Sharma is with University of Stuttgart.
  	E-mail: isha.sharma@visus.uni-stuttgart.com
  \item
  	Dieter Schmalstieg is with University of Stuttgart.
  	E-mail: dieter.schmalstieg@visus.uni-stuttgart.de.
}

\abstract{%
The complexity and scale of Volumetric and Simulation datasets for Scientific Visualization(SciVis) continue to grow. And the approaches and advantages of memory-efficient data formats and storage techniques for such datasets vary.

OpenVDB library and its VDB data format excels in memory efficiency through its hierarchical and dynamic tree structure, with active and inactive sub-trees for data storage. It is heavily used in current production renderers for both animation and rendering stages in VFX pipelines and photorealistic rendering of volumes and fluids. However, it still remains to be fully leveraged in SciVis where domains dealing with sparse scalar fields like porous media, time varying volumes such as tornado and weather simulation or high resolution simulation of Computational Fluid Dynamics present ample number of large challenging data sets.  

Goal of this paper is not only to explore the use of OpenVDB in SciVis but also to explore a level of detail(LOD) technique using 3D Gaussian particles approximating voxel regions. For rendering, we utilize NVIDIA OptiX library for ray marching through the  Gaussians particles. Data modeling using 3D Gaussians has been very popular lately due to success in stereoscopic image to 3D scene conversion using Gaussian Splatting as done by Kerbl et al.\cite{kerbl20233dgaussiansplattingrealtime} Gaussian approximation and mixture models aren't entirely new in SciVis as well, as was the goal of the paper by Jang et al.\cite{jang2006enhancing} Our work explores the integration with rendering software libraries like OpenVDB and OptiX to take advantage of their built-in memory compaction and hardware acceleration features, while also leveraging the performance capabilities of modern GPUs. Thus, we present a SciVis rendering approach that uses 3D Gaussians at varying LOD in a lossy scheme derived from VDB datasets, rather than focusing on photorealistic volume rendering. 

}

\keywords{OpenVDB, Ray marching, 3D Gaussians, OptiX ray tracing, Visual Computing}

\teaser{
  \centering
  \begin{minipage}[b]{0.329\linewidth}
    \includegraphics[width=\linewidth]{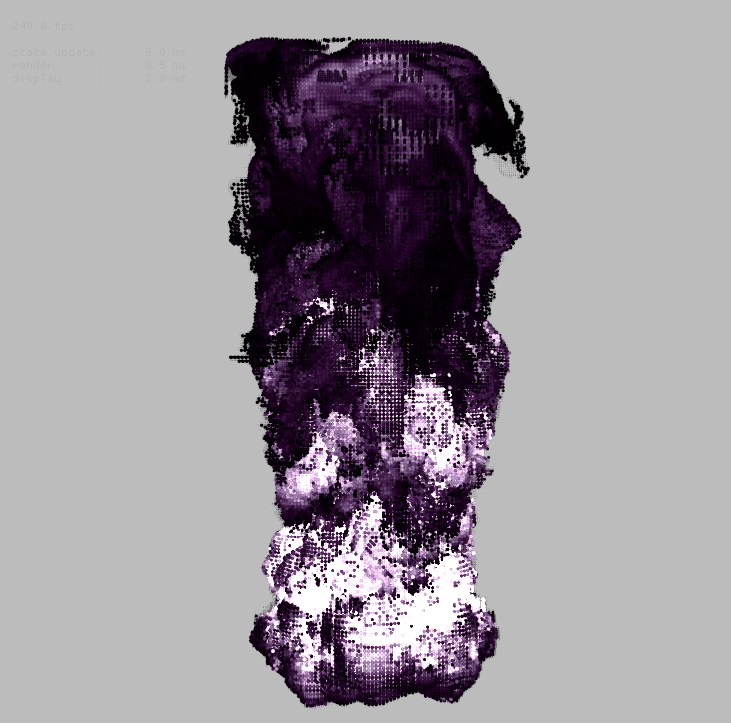}
  \end{minipage}
  \begin{minipage}[b]{0.329\linewidth}
    \includegraphics[width=\linewidth]{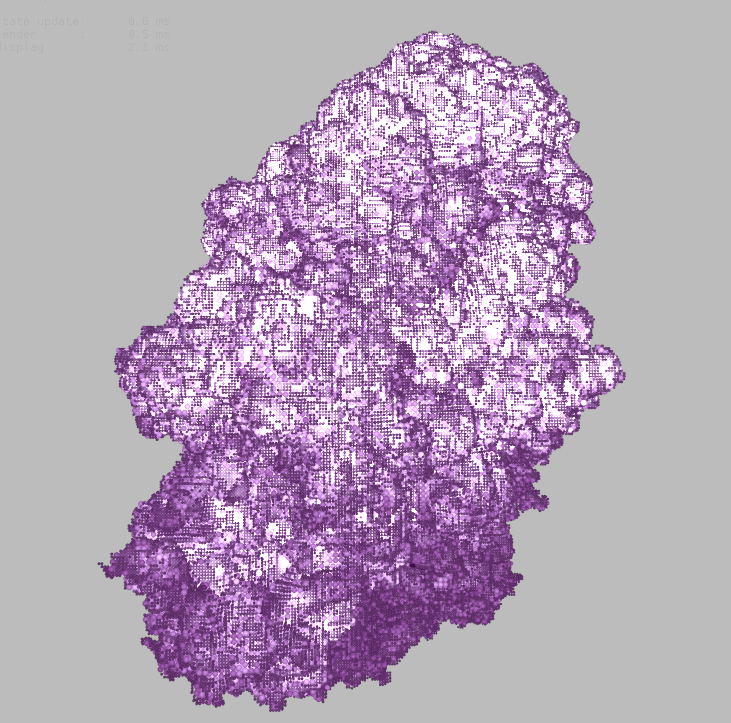}
  \end{minipage}
  \begin{minipage}[b]{0.329\linewidth}
    \includegraphics[width=\linewidth]{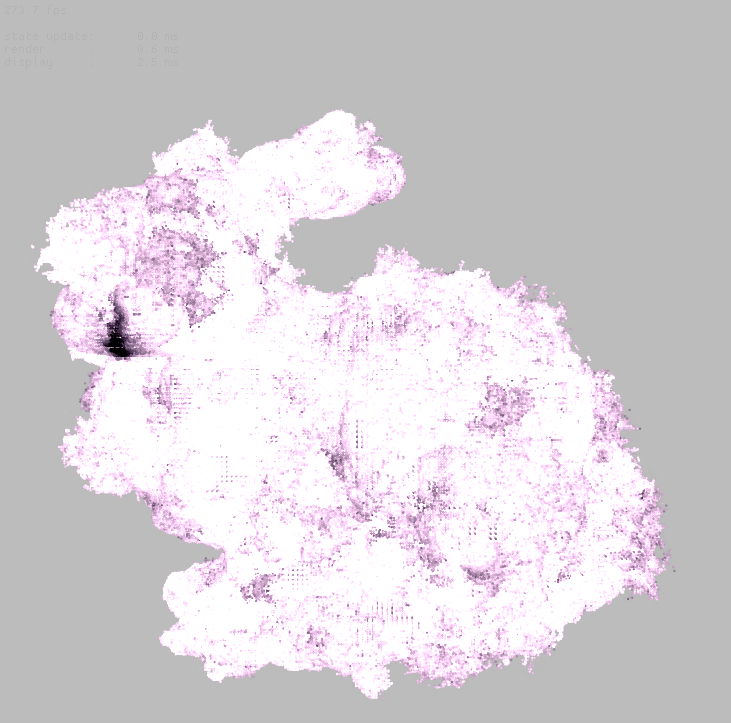}
  \end{minipage}
  \caption{OpenVDB datasets: fire, explosion and bunny\_cloud in 3D Gaussian approximation of the volumetric datasets}
  \label{fig:teaser}
}

\graphicspath{{figs/}{figures/}{pictures/}{images/}{./}} % where to search for the images

\usepackage{tabu}                      % only used for the table example
\usepackage{booktabs}                  % only used for the table example
\usepackage{lipsum}                    % used to generate placeholder text
\usepackage{mwe}                       % used to generate placeholder figures
\usepackage{textgreek} 
\usepackage{amsfonts}
\usepackage{makecell}
\usepackage{algorithm}
\usepackage{algorithmic}

\usepackage{mathptmx}                  % use matching math font
\usepackage{amsmath}
\usepackage[table,xcdraw]{xcolor}

\usepackage{fancyhdr}
\begin{document}

%%%%%%%%%%%%%%%%%%%%%%%%%%%%%%%%%%%%%%%%%%%%%%%%%%%%%%%%%%%%%%%%
%%%%%%%%%%%%%%%%%%%%%% START OF THE PAPER %%%%%%%%%%%%%%%%%%%%%%
%%%%%%%%%%%%%%%%%%%%%%%%%%%%%%%%%%%%%%%%%%%%%%%%%%%%%%%%%%%%%%%%

%% The ``\maketitle'' command must be the first command after the
%% ``\begin{document}'' command. It prepares and prints the title block.
%% the only exception to this rule is the \firstsection command
% \firstsection{Introduction}

\maketitle

\section{Introduction} %for journal use above \firstsection{..} instead
Apart from raw binary data, popular SciVis data formats can be grouped broadly into three categories. Formats like HDF5\cite{hdf5} and XDMF\cite{XDMF} offer advanced features such as compression, chunking and partial I/O, making them suitable for large structured grids. Formats commonly used in visualization pipelines, such as VTK\cite{VTK} and DICOM\cite{dicom}, tend to focus on dense memory storage and often configured without compression or hierarchical storage. Lastly, mesh-based formats like OBJ and STL provide simpler representations, but they offer limited to no data storage optimization capabilities. As the requirement for more efficient approaches to handling and optimizing large datasets arise, formats like OpenVDB present promising alternatives.
OpenVDB\cite{10.1145/2504435.2504454}, along with its GPU and neural variants(NanoVDB\cite{10.1145/3450623.3464653}, NeuralVDB \cite{kim2024neuralvdbhighresolutionsparsevolume} respectively) has been adopted for enabling memory efficient storage of sparse volumes and level sets across many domains, with the visual-effects industry being the biggest adopter and promoter of it. It is implemented in tools like Houdini, Blender, RenderMan and Arnold for rendering effects like  smoke, fire, explosions, clouds, atmospheric effects and volumetric lighting. While photorealistic volumes  and level sets for surface representation were its flagship use case, it is seeing more adoption outside of professional rendering tools, for e.g. in real-time and game-tech with NanoVDB and its fast GPU based ray marching for volumetric fog or fire. In Simulation industry for CFD\cite{FALTYNKOVA2025103822} and even many large scale multi-sensor data Visualization\cite{best2022resilient} use cases. However, the adoption of OpenVDB in SciVis use cases has been gradual, with its full potential yet to be fully realized.

VDB data format supports sparsity without wasting memory on empty regions, provides fast iterators  and accessors for voxel values and efficient filtering and sampling for data. This paper tries to explore the potential of VDB volumes for supporting SciVis use cases by analyzing the storage format features and using its Grid Tree nodes for implementing different levels-of-detail(LOD) for the datasets i.e., by using a reduced representation 3D Gaussian for each cluster of voxels.\footnote{This work is submitted to the IEEE for possible publication. Copyright may be transferred without notice, this version may no longer be accessible later.} We also implement a SciVis style forward ray marching renderer for the generated set of Gaussians with two transfer function color maps and provide resulting renderings of some of the OpenVDB's opensource volumetric datasets\cite{openvdb_download} along with a few level-sets for surface based representation. Ensuring compatibility with existing ray tracing frameworks, we use  NVIDIA OptiX\cite{parker2010optix}, to facilitate efficient ray casting and ray marching

%\textbf{please leave the copyright statement at the bottom-left of this first page untouched}.

\section{Goal}
Through implementation and use of a 3D Gaussian particle modeling and rendering scheme, the goal is to utilize VDB data format features for Scivis use cases such as: 
\begin{itemize}
\item Spatial clustering of data in leaf nodes of the OpenVDB Grid Tree(simply Grid from now on). Grid node properties such as "tiles" for exploring homogeneous voxel regions.
\item Grid leaf node bounding volumes in Index-space to OptiX mapping for efficient acceleration data structure construction for ray-box(Gaussian Axis-aligned Bounding-box) intersections.
\item Grid nodes regularity for LOD generation and adaption. 
\end{itemize}

And finally to carefully analyze the performance in terms of memory footprint of the Gaussian models. All of these features will be explained based on their context of implementation or use in sections below. 

\section{Previous works}
Most OpenVDB implementations are focused on rendering photorealistic volumes efficiently, with corresponding work integrated into software such as Blender or proprietary tools developed by NVIDIA and SideFX. Still, a few works have explored the use of OpenVDB in SciVis, medical visualization and large-scale simulation visualizations as well. 

Mayer et al.\cite{mayer2021visualization} present a method for visualizing human-scale blood flow simulations using Intel OSPRay Studio on the SuperMUC-NG supercomputer. They mapped the simulation data to memory-efficient VDB volumes, enabling interactive visualizations without extensive data preprocessing. 
Vizzo et al.\cite{vizzo2022vdbfusion} present VDBFusion, a versatile system for integrating range sensor data into truncated signed distance functions (TSDFs) using OpenVDB. Their approach leverages the effective application of OpenVDB mapping, enabling real-time processing of LiDAR data at 20 frames per second on a single-core CPU.
Bailey et al.\cite{bailey2015distributing} introduce a framework to integrate OpenVDB with OpenMPI for efficiently distributing liquid simulations across multiple processors. This approach addresses simulation of complex fluid dynamics in visual effects production. Walker et al.\cite{walker2022nanomap} introduce NanoMap, a GPU-accelerated mapping and simulation package that leverages OpenVDB and CUDA to efficiently process dense point clouds for robotic agents. This system significantly enhances real-time occupancy mapping and simulation capabilities, particularly on platforms with limited computational resources.

This section would be incomplete without mentioning the work by Borkiewicz et al.\cite{borkiewicz2017communicating}, who discuss the role of visualization in effectively communicating scientific concepts in an era where alternative facts are prevalent. They emphasize the importance of integrating scientific accuracy with compelling visual narratives to enhance public understanding and engagement.
Regarding LOD from Gaussian data, Seo et al. \cite{seo2024flod} introduced a Flexible Level of Detail (FLoD) technique, however for 3D Gaussian Splatting. Our work is inspired by the recent widespread adoption of OpenVDB across other domains and it is novel in its approach to generating Gaussian models of varying LOD from VDB data.

\section{Data Generation Model}
The Gaussian model is generated from data stored in the VDB Grid nodes and hence a primary understanding of the layout is essential before re-purposing the data. The Grid structure is infact a deciding factor for clustering of data during LOD improvements in our scheme. 

\subsection{Grid Layout}
This section offers concise explanation of VDB format features which are necessary for understanding the data generation model later and might be challenging to conceptualize from OpenVDB documentation\cite{openvdb_documentation} or their scientific paper.

The data representation of voxels is an axis-aligned and regularly-spaced Grid structure. 
The Root node spans the entirety of the dataset which is subdivided into multiple top-level nodes that are cubical sub-regions and have a power of 2 along each dimension.
These top-level nodes are identified with  their unique origins i.e. their location in 3D space. Their quantity is only constrained by the need to cover the entire data. We have observed that they exist in a non-overlapping manner. This is an important distinction from other widely used Gaussian models, which is discussed in the data generation subsection later.

The most common grid layout is the 5-4-3 variant and we will use this layout for the explanation, with all rendering samples also adhering to the same layout. In this variant each top-level node or "5-level" node is $2^5 \times 2^5 \times 2^5$ in resolution, i.e. it contains 32768 total children. 
Each of these children is an intermediate-level or "4-level" node containing $2^4 \times 2^4 \times 2^4$ i.e. 4096 children. 
Subsequently, each of those children can be further subdivided into $2^3 \times 2^3 \times 2^3$ final voxels. The "3-level" nodes are also called the leaf nodes and they contain 512 voxels of data each. 

The depth of the tree is fixed and the number of top-level nodes is adaptive to span the whole dataset, hence making node/sub-region access deterministic.
Each node at top-level and intermediate-level has a fixed number of children, this is similar to many commonly used tree data structures that have a fixed number of children at any level. This feature is crucial to deterministic allocation of active and inactive regions of a sparse dataset densely in memory.

If the nodes were densely packed in memory it would provide no benefit over a standard oct-tree or any other hierarchical tree structure, however in OpenVDB not all voxels are stored, as it supports sparsity by marking a voxel as active or inactive. $4096 \times 4096 \times 4096$ children of one top-level node can alone consume over 128GB of data with merely a half-floating point precision values. 

To support selective sparse data storage there are two kinds of masks for nodes at each level
\begin{itemize}
    \item Child mask: 1 bit is stored per child in every node at each level to mark, whether the sub-tree of that child is active or not. For e.g., for one 4-level node we need 1 bit per 3-node child i.e. 4096 bits in total. This value can be stored in 64 64-bit integers for this entire node.   
    \item Value mask: It has the same size of 1 bit per child in a node of a particular level however its value suggests whether same data value can the used for the entire sub-tree of the child. Such subregions that omit creation of a sub-tree and store rather a single value for the entire sub-tree are called "Tiles".
\end{itemize}

\subsection{3D Volumetric Gaussians generation}
Converting the grid tree into 3D Gaussians can be achieved in various ways and for the purposes of this paper, we have chosen a non-overlapping model i.e., every gaussian is non-intersecting with other Gaussians and spans a voxel sub-region in a reduced-fidelity representation due to spherical or ellipsoidal representation of voxelated volumetric regions.
We will discuss also other possible models for providing a complete picture of the idea.

\begin{figure}[tbp]
  \centering
  \begin{subfigure}[b]{0.48\columnwidth}
    \centering
    \includegraphics[width=\textwidth, alt={Big letter A on a gray background.}]{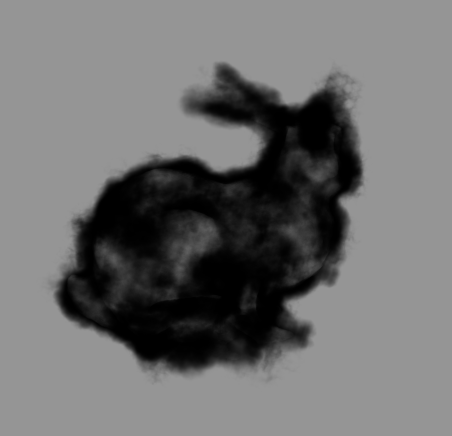}
    \caption{Photorealistic volume rendering}
    \label{fig:ex_subfigs_a}
  \end{subfigure}%
  \hspace{0.02\columnwidth}% Reduced space between subfigures
  \begin{subfigure}[b]{0.48\columnwidth}
    \centering
    \includegraphics[width=\textwidth, alt={Big letter B on a gray background.}]{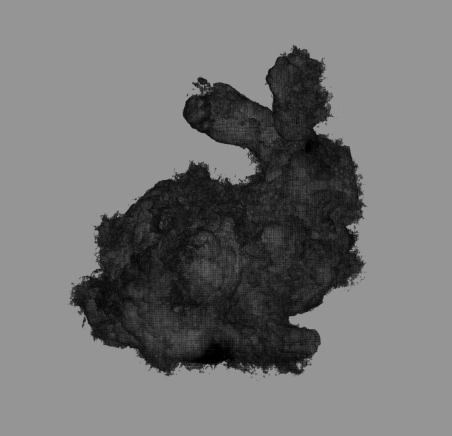}
    \caption{3D Gaussians particles from leaf nodes}
    \label{fig:ex_subfigs_b}
  \end{subfigure}%

  \subfigsCaption{3D Gaussians approximating VDB leaf nodes in the bunny\_cloud dataset.}
  \label{fig:ex_subfigs}
\end{figure}

\subsubsection{Leaf nodes Conversion}
The simplest method to convert voxels into 3D Gaussians is through the OpenVDB API, where all leaf nodes are retrieved and transformed into Gaussians at various resolutions based on the LOD. We derive the following three properties for every Gaussian in our model:
\begin{enumerate}
    \item position: a well-defined center in world space, derived from the centroid of a voxel region or bounding box
    \item opacity: density value to reflect the average scalar value of the region, preserve the local material  characteristics
    \item covariance vector: diagonal vector of the 3x3 covariance matrix representing its spatial extent in each axis, since our Gaussians are axis-aligned with the three axes. Or a single scalar value in case of isotropic Gaussians. This enables the representation of both spherical and ellipsoidal shaped Gaussians.
\end{enumerate}

Before directly using the leaf nodes, it is essential to verify the \textbf{Tiled} data storage, as this ensures that no data is omitted. Some leaf nodes may be excluded from creation, if they have a fixed value for every voxel. The value is then stored with a value mask in their intermediate-level node parent and the leaf simply ceases to exist. Although tiles provide the easiest way to detect homogeneous data to convert them into a single Gaussian, in this case they create an overhead of traversing the Grid Tree, as opposed to simply fetching nodes of a particular level which is much easier. Tiles can exist on both intermediate and top-level nodes.
    
\textbf{Leaf Nodes} themselves can be dense or sparse i.e., they're either fully packed with values or partially empty. But their resolution is well defined i.e. 512 voxels, hence we can perform parallel processing on them. 
    Intel Thread Building Blocks(TBB or oneAPI TBB)\cite{intel_onetbb} is a required dependency of OpenVDB for CPU side thread parallelism for tasks such as iterating over leaf nodes, building or transforming grids and performing most voxel operations like filtering or resampling. TBB is not part of the OpenVDB library itself but it depends on it as default backend for multi-threading.
    
    On a high level, each thread can process one leaf node at a time, collect Gaussian data such as position, opacity and covariance diagonal vector in thread-local buffers and finally write to a global array of generated Gaussians using a mutex for safety. Each thread has its own: 
    \begin{itemize}
        \item \texttt{thread\_gaussian\_positions}: List of 3D centers (positions of Gaussians).
        \item \texttt{thread\_gaussian\_opacities}: Scalar opacities.
        \item \texttt{thread\_gaussian\_covariances}: Vector of diagonal elements of a 3×3 matrix (as flat array) encoding variances in X, Y and Z axis respectively. Since the Gaussians are axis-aligned for fast approximations currently, the rotation is not encoded. 
        \item \texttt{localLeafData}: Metadata for setting up Axis-aligned Bounding-boxes(AABB) for Gaussians for creating an Optix GAS(Geometry Acceleration Structure), point offsets and count in the global buffer if more than one gaussian is stored per AABB. For simplicity, in our renderings we used one Gaussian per AABB. Hence eliminating the need for offsets and count. 
    \end{itemize}
    
 \subsubsection{LOD implementation}  
 LOD of the Gaussians approximations is based on three important factors:
\begin{itemize}
    \item The specific region of the grid used for model construction(for e.g. only leaf nodes or entire Tree),
    \item Compliance to the grid space partitioning and
    \item The clustering method and its refinement.
\end{itemize}
Since the leaves are essentially non-overlapping, due to the strictly non-overlapping top-level nodes and their symmetric geometry (every node is a cubic regions of $2^n$ resolution, where n=5,4,3), our method utilizes three overarching refinement levels, each introducing varying degrees of lossiness during rendering. This is primarily because, so far, we have only experimented with non-overlapping Gaussians. For completeness, we also present conceptual ideas for supporting overlapping Gaussians, along with a formulation of the Gaussian Mixture Model, which are discussed in the sections on the Inter-leaf node Gaussian model and Grid Tree mirroring below.
\begin{itemize}

\item{Spatial Clustering and Averaging: }

This method focuses on speed and maintaining spatial coherence with the original structured data. The OpenVDB representation of space can be either index-based or in world space. Certain operations are performed in Index Space to allow efficient integer-based grid operations. Bounding boxes of nodes of the Grid Tree are defined in Index Space and characterized by their minimum and maximum integer  coordinates. By applying the Grid transform we can easily calculate their world coordinates, although the API requires us to obtain the translation part separately in addition to the transformation matrix. This was slightly unique in our experience of working with rendering libraries. 
The world coordinates of any point can be easily calculated once you have the full transformation matrix(\emph{M}) given by: $\mathit{p}_{\text{world}} = M \cdot \mathit{p}_{\text{index}}$  

We have thus the following three defined LOD which is used in our renderer:

\emph{Low LOD: }
This is approximation of an entire leaf node with a single Gaussian. The leaf nodes already provide a good starting point for data clustering because of their limited and consistent size(512 voxels in total) and can be considered as default ready-made clusters for the lowest LOD level. It would be sub-optimal to not make use of these spatial "clusters", we have for both dense and sparse leaf nodes the following:

\begin{itemize} 
    \item Sparse leaf nodes:
    We calculate the axis-aligned bounding box of a sparse leaf node by traversing all active voxels in the node and iteratively expanding the bounding box which is initialized to the dimensions of the starting voxel. If no active voxels are found, the function exits early.
    The minimum and maximum coordinates of the node's "active" bounding box are calculated in VDB index space. The centroid of the active voxel region transformed into world-space serves as the Gaussian position. The bounding box size is used to derive the Gaussian's radii, scaled according to the voxel dimensions. 
    \[
\boldsymbol{\sigma} = \frac{1}{2} \cdot (bboxMax - bboxMin) \cdot voxelSize
\] 
where $\sigma$ and \texttt{voxelSize} is the vector of 3 floating point values of radii in each axis and voxel size respectively. 
    The algorithm then averages the opacity values of all active voxels to estimate the overall opacity, optionally scaling the result to account for the Gaussian's spatial extent. A diagonal covariance matrix is constructed from the radii, defining an axis-aligned ellipsoid that coarsely approximate the region. Finally, the Gaussian's parameters: position, opacity and covariance are stored in output buffers with the corresponding metadata for later use during rendering or export. These sparse leaf Gaussians are mostly ellipsoids.
    \item Dense leaf nodes:  Due to consistent structure of dense nodes and defined voxel sizes of the Grid, the center of the Gaussian can be easily calculated in world coordinates through its Index space bounding boxes, making these Gaussians mostly spheres, specifically if the \texttt{voxelSize} is isotropic in all three dimensions making the Gaussian's radii isotropic.  
    
\end{itemize}

\emph{Medium LOD: }
The medium LOD approximation is implemented through two structurally distinct algorithms, depending on whether the target leaf node is dense or sparse. For dense leaf nodes, we apply a uniform spatial partitioning scheme, subdividing the voxel region into non-overlapping 4×4×4 blocks (64 voxels each). For every block, we compute the mean scalar value to determine opacity, place the Gaussian center at the geometric center of the block in index space (mapped to world coordinates via the grid transform) and assign axis-aligned covariance values proportional to the \texttt{voxelSize}. This produces a coarse but spatially consistent Gaussian layout that reduces detail while preserving volumetric distribution.

For non-dense leaves, where active voxels are irregularly distributed, a localized block-based clustering strategy is used. The algorithm traverses the leaf and attempts to form 2×2×2 blocks of adjacent active voxels. If a complete block is found, a Gaussian is emitted with its center located at the block’s centroid and a covariance corresponding to \texttt{voxelSize} scaled radii. Otherwise, a fallback is used where isolated active voxels are each represented by an individual Gaussian centered at the voxel midpoint, with half-voxel radii. This hybrid scheme allows spatial adaptivity: compact clusters are merged into a single Gaussian for efficiency, while isolated voxels are retained individually to preserve structural detail. Both methods store the resulting Gaussian parameters—including position, average opacity, covariance and metadata for OptiX rendering.

\emph{High LOD: }
The highest LOD in our representation retains local detail as much as possible. 
It operates the same way as Medium LOD, however the subdivision of dense leaf nodes happens into 2×2×2 blocks, effectively aggregating every 8 adjacent voxels into a single Gaussian. The scalar values of the 8 voxels are averaged to compute the Gaussian's opacity, while the block center in index space (offset by \texttt{voxelSize}) is transformed into world space to serve as the Gaussian's position. The covariance is axis-aligned and corresponds to the full voxel size along each axis. This results in a dense tiling of Gaussians that approximates the field at near-voxel resolution while reducing primitive count through regular grouping.

Non-dense leaf nodes in this setting are treated at the per-voxel level, where each active voxel is individually converted into a Gaussian primitive. The voxel center (offset by half the \texttt{voxelSize}) is used as the Gaussian mean and the opacity is taken directly from the voxel value. The Gaussian's radii are set to half the voxel size, effectively treating the voxel as a minimal, spatially localized Gaussian. This produces a one-to-one mapping between active voxels and Gaussians, maximizing detail retention in sparse regions. 
While this approach of approximating one Gaussian per voxel may seem simplistic, it serves as a starting point for exploring a feature-preserving optimal clustering strategy. Due to the uneven spread of active voxels in sparse leaves it is non-trivial to select the clustering strategy that provides the right coverage for all, at times widely scattered voxels.

% This is the setting that gave us better results at present so we have produced all renderings from this. 
% Dense and non-dense nodes are to be handled separately  
 
\item Variance based Intra-leaf Voxels clustering:
We implemented a clustering method that adaptively fits anisotropic 3D Gaussians to sparse volumetric data, by recursively splitting voxel regions based on variance until local homogeneity is achieved. 

    \begin{enumerate}
    \item Compute the centroid and average value of the active voxels in a leaf node.

    \item Compute the population variance, which defined how much the voxel values differ from the average.
Mathematically, it is given by:
$$
\text{variance} = \frac{1}{N} \sum_{i=1}^{N} \left\| x_i - \mu \right\|^2
$$
    where \( N \) is the number of active voxels, \( x_i \) is the world-space position of the \( i \)-th voxel and \( \mu \) is the centroid.

    \item If the variance is below our threshold, we emit a Gaussian for the block of voxels.

    \item If the variance is large, the block is split into two along the axis with the most variance and each half is processed recursively.
\end{enumerate}
We found that this method was not as effective for non-overlapping clustering compared to the spatial clustering approach described earlier. Therefore, for the results, we have used only the spatial clustering method.

\item{Inter-leaf node clustering: }
In this variant the leaf nodes would be processed as described in the  section on Spatial Clustering, however sets of neighboring blocks would be assessed in \emph{a second pass} for finding cross-leaf homogeneous regions for merging and forming a bigger Gaussian. 
Since resolution or LOD is mutually exclusive of this process, it can apply to all the three methods of refinement described above. 

For this technique, we must store intermediate LOD refinement results as if they were individual leaf nodes themselves and run spatial variance estimation on these results sequentially in groups of 26 for example: considering each 2x2x2 node will have 26 neighbours. These neighbors can be fetched based on their bounding box indices for "faces", "edges" and "corners" neighbors.

If the computed variance is below a predefined threshold, indicating that the region is sufficiently uniform and we can tightly clustered it further and replace the entire group with a single merged Gaussian. The new bounding box spans from the minimum of all lower bound to the maximum of all the upper bounds of the bounding boxes of regions processed. 
With \emph{C} as the corner coordinates we have: $\mathbf{bbox}_{\min} = \min_i \left( \mathbf{c}_i^{\min} \right)$, 
$\mathbf{bbox}_{\max} = \max_i \left( \mathbf{c}_i^{\max} \right)$. The Gaussian position is the new merged centroid and the axis-aligned radii and opacity is calculated much like the variants already described.
\item{Grid Tree mirroring: }
 Gaussian Model generation that can mimic the hierarchy of the OpenVDB Grid Tree is the essentially the Gaussian Mixture Model that can provide the highest data fidelity with least lossiness(fewer numbers of rounded Gaussian corners approximating voxel data). 
To produce a nested set of overlapping and encompassing Gaussians across multiple scales, all nodes of the Tree have to be processed. This is equivalent to flattening the entire VDB Grid Tree once and hence much more complicated to implement efficiently than the above described leaf node based spatial clustering alternatives. VDB Grid is although a multi-resolution partitioning of space, where each node covers a fixed, axis-aligned region of the volume, the values are only stored in leaf nodes and without having an estimate of data values possessed by all of the children of an intermediate or top-level node, an approximate value for the parent node in consideration cannot be provided.

Although this allows to progressively construct a coarse-to-fine Gaussian representation, where higher-level Gaussians (in root or internal nodes) cover the left-over regions from rounded corners of the refined Gaussian approximations of fine detail. Flattening of the Tree will only grow more tedious with the increasing size of the dataset, hence we are still investigating the benefit of this technique over the initial processing required. 
Unlike strict partitioning approaches, this overlapping Gaussians method shall provide compatibility with standard Gaussian mixture rendering pipelines, where overlapping regions contribute proportionally to the final density and color.
\end{itemize}

A note on \textbf{CPU vs GPU data generation}(OpenVDB vs NanoVDB): OptiX requires device-side arrays for Gaussian particles and NanoVDB's particle data grids seem ideal for this purpose. We attempted loading .vdb files as NanoVDB float grid and process it on the GPU for Gaussian particles creation. We also tried using NanoVDB particle grids allocation of our Gaussians for ray marching using OptiX. However, the lack of device-side iterators or accessors creates a significant bottleneck, especially when random access is required, as in ray-BVH or ray-particle intersections in OptiX.
\begin{itemize}
    \item 
    NanoVDB does not support traversing the Grid Tree on the device, which is crucial for generating Gaussians. As a result, it cannot be used for 3D Gaussian generation in any case. Therefore, all Gaussian generation was moved to OpenVDB for this work.
\item
    NanoVDB provides built-in support for calculating transmittance using the HDDA\cite{10.1145/2614106.2614136} algorithm for float grid volumes. However, this method is not applicable to particles with varying sizes that might be overlapping.
\item
    While NanoVDB is the GPU-compatible version of OpenVDB, its unit tests and documentation cite particle data support, OptiX SDK only provides implementations for float grid volumes. As of now, we are unaware of any working support for NanoVDB particle data grids with the OptiX ray-tracing engine.
\end{itemize}
OptiX ray-hit primitive indices are returned during closest-hit tests. For transmission and color accumulation along the ray we require accessing and processing primitive properties during the ray marching. For float grids, this is done by NanoVDB's built-in HDDA transmittance functions. It processes the accumulation internally and this technique only works with regular float grid volumes, for overlapping and variable-size particle data with custom ray tracing pipeline this is invalid. Efficient traversal and data retrieval mechanisms along the ray are required for volumetric particle ray marching and this seemed deficient using NanoVDB.

Mechanisms like the NodeManager and iterators are limited to CPU-side traversal. It is impractical to traverse data on the CPU and then update GPU memory each time. After extensive exploration and testing of the NanoVDB library, we determined that its features were insufficient for our use case. Therefore, we decided not to use NanoVDB in favor of OpenVDB, which better supports the requirements for our GPU-based rendering pipeline.

\section{Gaussians Renderer}
The 3D volumetric Gaussians generated from OpenVDB node data necessitate a specialized renderer capable of traversing from one Gaussian to the next along the camera ray direction, from the entry point to the exit point of the entire volumetric dataset.
We have hence created an interactive viewer and renderer in C++ with GLFW as the library for supporting real-time user-interaction via keyboard/mouse input and a trackball-style camera with zoom, pan and rotation for camera motion. We use NVIDIA's OptiX ray tracing framework for GPU-accelerated ray marching of the 3D volumetric Gaussian particles. For the hardware setup we have used NVIDIA GeForce RTX 4090 Laptop GPU with Intel Core i9-14900HX. Since OpenVDB requires CPU-side multi-threading for voxels and nodes processing and OptiX is entirely written for device side operations, having a good combination of those two processors is essential. 
Before we can use hardware acceleration, all of the data generated from the Gaussian generation should be copied over as separate buffers to the device. 

\emph{Device copy of data buffers: } For initializing device-side buffers with the 3D Gaussians data, CUDA operations such as \texttt{cudaMalloc} and \texttt{cudaMemcpy} for host-device data transfer are required. Every Gaussian property such as opacity(scalar), world position(3D vector) and covariance(3D vectors or scalar based on whether Gaussians are spheres or ellipsoids) has its own buffer which is allocated contiguously on the device side for all Gaussians. The device side pointers to these buffers are part of a parent class \texttt{GaussianModel}, whose single instance or object is added to the OptiX Shader Binding Table for enabling access to these buffer. Once the buffers have been copied from host to device they can be safely deleted on host's side.

Since this paper does not concern with explaining the OptiX's \emph{modus operandi}, we encourage reading their documentation\cite{nvidia_optix8} for more detailed information on it. However we provide sufficient explanation of technical concepts and  different stages of the rendering pipeline required for understanding our implementation of the Gaussians renderer.

\subsection{Acceleration Data Structures} 
\begin{itemize}
    \item GAS(Geometry Acceleration Structure): This Acceleration construct is responsible for efficient ray-geometry intersection tests by organizing primitives such as AABBs, triangles or any other custom primitives in a way that minimizes the number of intersections needed during rendering. 
    GAS operates in Index Space. Hence this is where index space coordinates from OpenVDB can be directly mapped. This can be exploited for performance much in the same way as OpenVDB intended to with volume accesses on the data representation side. 
    In our case we build the GAS with a CUDA device side pointer of an array of axis-aligned bounding boxes, using the minimum and  maximum coordinates of the Gaussian "particles" AABBs in Index Space. These AABBs are the extents of each Gaussian generated using different LOD levels during the generation pipeline.
    There are many build and memory computation settings available in the OptiX framework, however given the limited time for testing this setup we tried only \texttt{OPTIX\_BUILD\_FLAG\_ALLOW\_COMPACTION} and with \texttt{OPTIX\_BUILD\_INPUT\_TYPE\_CUSTOM\_PRIMITIVES} as the build input. 
    \item IAS(Instance Acceleration Structure): It functions similar to a \emph{Scene Graph} and stores instances of GAS objects, each with their own transforms. In our case the transform used is the \emph{OpenVDB Grid transform} for our GAS instance. 
    
    It is therefore a key advantage of this system, to be able to perform direct and no-conversion mapping between OpenVDB grid instances to Optix IAS. This enables working with OpenVDB index spaces in OptiX. In addition it can also store visibility masks. Every ray and every OptiX instance (geometry) has a visibility mask. When a ray is traced, only geometry instances whose mask overlaps (bitwise AND) with the ray’s visibility mask will be considered for intersection. The Gaussians AABBs are all set to visible for the primary and continuation rays. This setting can be used for future performance upgrade of such a  system by skipping certain Gaussian groups entirely. Each AABB acts as a bounding volume for a single Gaussian. During ray traversal: OptiX checks visibility masks at the instance level (i.e., per AABB). If a ray passes the mask test and intersects the AABB, the intersection shader runs and checks whether the ray actually intersects the Gaussian inside.
\end{itemize}

\emph{Shader-binding Table(SBT): }
SBT holds all per-primitive and per-material data required by OptiX shaders. 
 Although the Gaussian properties buffers are allocated on the device, they still need to be explicitly passed to the OptiX pipeline during ray marching for efficient access without rebuilding the SBT.  Even though the SBT lives on the GPU, the GUI (e.g., GLFW key callbacks) runs on the CPU and shared control of the material parameters such as opacity, in real-time is required by interactive applications.
 A HitGroupRecord is to be initialized for the Gaussian particles rendering. It is a structure that specifies the shaders associated with a particular type of primitive(in our case a Gaussian) that the ray might intersect during the ray tracing process. A pointer to the \texttt{GaussianModel} parent class object containing device side pointers for each Gaussian property array is provided to this record.

\subsection{Ray Tracing Pipeline}
OptiX allows the use of distinct hit programs tailored to different types of scene geometry. In the following, we focus exclusively on the hit programs relevant to 3D Gaussians that act as volumetric particles.
We utilize the \texttt{\_\_intersection\_\_} and \texttt{\_\_closesthit\_\_radiance\_} Programs workflow with OptiX for ray tracing. The \texttt{\_\_raygen\_\_} program for shooting primary rays is the same as any other standard volume rendering with OptiX.

\subsubsection{Intersection} We implemented our intersection program to determine if a ray intersects any provided AABBs. Using the primitive index provided by OptiX, the program determines which subset of the Gaussian dataset corresponds to this leaf node. Currently we have one Gaussian per AABB, the offset and count metadata are used to iterate correct Gaussians subset in case of multiple Gaussian entries per AABB. For each Gaussian, it retrieves the center and covariance and the ray is transformed into the Gaussian's local space using the inverse covariance matrix. The quadratic equation(provided in the appendix) is solved to determine the intersection points of entry and exit with the ellipsoid or sphere, along the ray.  By leveraging this AABB selection and iterating only through relevant Gaussians, we avoid many a ray-primitive intersections tests.
We exit the intersection shader early upon finding intersection and populate payloads for later use in the closest-hit programs.

Each Gaussian is enclosed in its own AABB. When a ray intersects a node, we:
\begin{itemize}
    \item Perform the ray-box intersection with the world AABB. 
    \item Use the primitive index to retrieve the Gaussian AABB that is intersected to ensure that the ray is within the bounding volume of the Gaussian. This will allow us to operate only on the Gaussian(s) that is in the ray's path. 
    \item Iterate through Gaussian(s), fetch position (\( \boldsymbol{\mu} \)), inverse covariance (\( \Sigma^{-1} \)) and opacity (\( \alpha \)).
    \item Solve the quadratic equation provided in the appendix for the intersection test.
    \item On intersection, store entry (\( t_0 \)) and exit (\( t_1 \)) distances in the payload.
\end{itemize}

\subsubsection{Rendering Loop}

Our renderer performs ray marching through Gaussians using a closest-hit-based pipeline implemented in OptiX. Unlike traditional scalar-field-based volume rendering, we treat each 3D Gaussian as a particle that represents a Gaussian distribution of local opacity values. This approach is essential for calculating the density at each sample point along the ray, which is then accumulated over a custom line integral to provide the accumulated opacity values.
\begin{itemize}
    \item {Ray Marching and Integration: }
Our closest-hit shader performs fixed-step ray marching through the Gaussian density field. The ray is initialized in world-space and its direction is provided from the raygen program. The individual Gaussian's entry and exit points come via the intersection shader as OptiX payloads. We calculate the density at a defined number of sample points between the entry and exit as we step along the ray using predefined step size based on the distance between the entry and exit point. 

\item{Density Evaluation:}
To evaluate the Gaussian density at a point \( \mathbf{x}_k \), we begin by computing the difference between the sample point and the center of the Gaussian i.e., $\boldsymbol{\delta} = \mathbf{x}_k - \boldsymbol{\mu}$ to compute the Mahalanobis distance\cite{holbert2022outlier} squared:
\[
    D^2 = \boldsymbol{\delta}^\top \Sigma^{-1} \boldsymbol{\delta}
\]
where \( \Sigma^{-1} \) is the inverse of the covariance matrix of the Gaussian and \( \boldsymbol{\mu} \) is the mean (center) of the Gaussian distribution. This distance quantifies how far the sample point is from the mean.
The Gaussian density at the sample point \( \mathbf{x}_k \) is then evaluated as:
\[
    \rho_k = \exp\left( -\frac{1}{2} D^2 \right)
    = \exp\left( -\frac{1}{2} (\mathbf{x}_k - \boldsymbol{\mu})^\top \Sigma^{-1} (\mathbf{x}_k - \boldsymbol{\mu}) \right)
\]

This density value \( \rho_k \) is used in the radiance accumulation to determine the contribution of the Gaussian at that point along the ray.

Next, we adjust the absorption based on the volume\_factor which accounts for the spread or volume of the Gaussian distribution. 
\[
    \text{volume\_factor} = {1}/{\det(\Sigma)}
\]

Finally, the absorption(A\textsubscript{k}) at each sample point is calculated as:
\[
    \text{absorption} = \rho_k \cdot \Delta t \cdot volume\_factor
\]
where: \( \rho_k \) is the Gaussian density at the sample point,
\( \Delta t \) is the step size, which represents the distance between two consecutive sample points along the ray and 
\( \det(\Sigma) \) is the determinant of the covariance matrix \( \Sigma \).
The Beer-Lambert law\cite{scratchapixel_volume_rendering} relates to the absorption of light as it passes through a medium and the principles behind it are applied in the absorption and transmittance calculations here.
\paragraph{Radiance Accumulation:}
At each sample point, the radiance is accumulated by the formula:
\[
    \mathbf{C} \mathrel{+}= T_k \cdot \mathbf{c}(\sigma_k) \cdot A_k
\]
where:
\( \mathbf{C} \) is the accumulated color, 
\( T_k \) is the transmittance at the current sample point, 
\( \mathbf{c}(\sigma_k) \) is the scattering color derived from the opacity \( \sigma_k \) (potentially via a transfer function) and 
\( A_k \) is the absorption at the current sample point.

The transmittance is updated using the exponential decay based on the absorption \( A_k \):

\[
    T_{k+1} = T_k \cdot \exp(-A_k)
\]

where \( T_{k+1} \) is the updated transmittance after the current sample point.

\item{Ray Continuation:}
After marching within the Gaussian, the ray origin is advanced to \( \mathbf{o} + \mathbf{d} \cdot (t_1 - t_0) \), where:
\( \mathbf{o} \) is the current ray origin,
\( \mathbf{d} \) is the ray direction and 
\( t_1 \) and \( t_0 \) are the entry and exit points of the Gaussian along the ray, respectively.

A recursive call to \texttt{optixTrace()} is then made to trace the next Gaussian along the ray.
This process continues until the transmittance falls below a threshold or the maximum recursion depth is reached. A pseudo code of the recursive ray tracing algorithm is provided below:
\end{itemize}
\begin{algorithm}
\begin{algorithmic}[1]
\STATE \textbf{recursiveRayTrace(ray, t\_min, t\_max, max\_depth)}
\STATE \quad Init color, transmittance = 1.0, recursion\_depth = 0
\WHILE{recursion\_depth < max\_depth}
    \STATE \quad t0, t1 = t\_min, t\_max
    \IF{not intersectWorldAABB(ray, worldBBox, t0, t1)}
        \STATE \quad return background\_color
    \ENDIF
    \FOR{each Gaussian in AABB}
        \IF{rayIntersectsGaussian(ray, Gaussian, t0, t1)}
\STATE \quad \( \text{density} = \text{computeGaussianDensity}( \text{ray} \times t_0, \) \\
\quad \( \text{Gaussian\_pos}, \text{inverse}(\text{covariance}), \alpha )
\)
            \STATE \quad \( \text{volume\_factor} = 1.0 / {\text{determinant}(covariance)} \)
            \STATE \quad \( \text{absorption} = \text{density} * (t1 - t0) * \text{volume\_factor} \)
    
\STATE \quad \( \text{color} += \text{transmittance} \times \text{scattering\_color} \times \text{absorption},  \text{transmittance} *= \exp(-\text{absorption}) \)

            \STATE \quad break
        \ENDIF
    \ENDFOR

    \STATE \quad \( \text{ray\_orig} += \text{ray\_dir} \times (t1 - t0) \) 
    \STATE \quad \text{optixTrace}(\text{payload})

    \IF{recursion\_depth > 10}
        \STATE \quad break
    \ENDIF
\ENDWHILE

\STATE \quad \( \text{color} += \text{transmittance} \times \text{payload.result} \)
\STATE \quad return color
\end{algorithmic}
\end{algorithm}

\emph{Limitations: }Currently, radiance integration only occurs within the closest intersected Gaussian. Because we do not use \texttt{\_\_anyhit\_\_} shaders, overlapping Gaussians are ignored, leading to incorrect transparency and color accumulation. Attempts to propagate state via recursive \texttt{optixTrace} calls have been limited by OptiX’s 8-payload register constraint. We plan to adopt \texttt{\_\_anyhit\_\_} or global buffer accumulation in future versions and are evaluating the best path forward for correct multi-Gaussian integration.

 We are not tracing shadow rays or scattering rays considering the supported use case of this renderer is strictly SciVis applications. All optixTrace calls are made with RAY\_TYPE\_RADIANCE. The current renderer is purely forward-ray-marching for color and opacity updates only. 

\subsubsection{Transfer functions}
\begin{figure}[!t]  
  \centering
  \setlength{\tabcolsep}{2pt}
  \vspace*{-\topskip}  
  \begin{tabular}{cc}
    \includegraphics[width=0.25\textwidth]{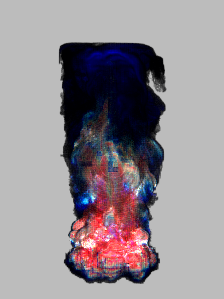} & 
    \includegraphics[width=0.25\textwidth]{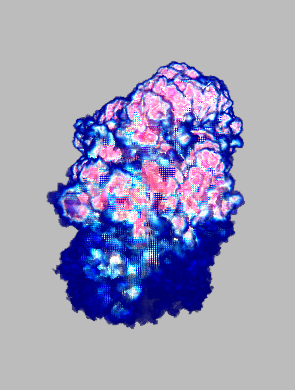} \\
    \includegraphics[width=0.25\textwidth]{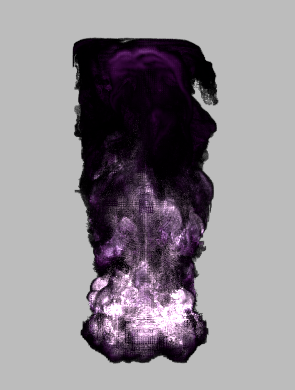} & 
    \includegraphics[width=0.25\textwidth]{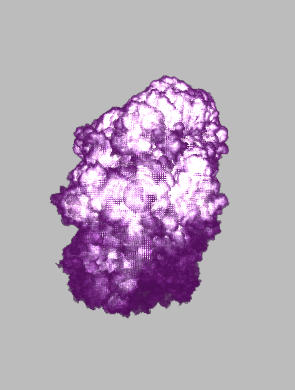} \\
    \includegraphics[width=0.25\textwidth]{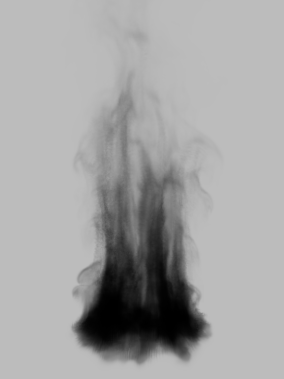} & 
    \includegraphics[width=0.25\textwidth]{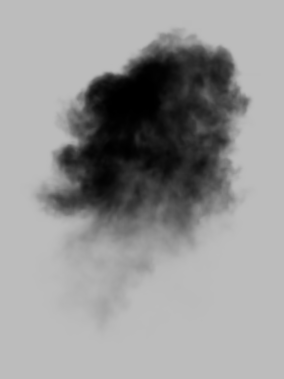} \\
  \end{tabular}

  \caption{OpenVDB datasets \emph{fire}(left) and \emph{explosion}(right) in two different transfer functions. At the top row we have jet and in the middle viridis. At the bottom row we have photorealistic rendering with NanoVDB HDDA transmittance for comparison}
  \label{fig:collage_part2}
\end{figure}

Unlike other Gaussian splatting based implementations, we do not have color baked in our Gaussians. Hence, our Gaussians renderer naturally supports transfer functions, similar to traditional volume rendering. Scalar attributes like opacity can be assigned color values through colormap lookups(transfer function maps). Currently, we use static color maps implemented on in the shader sources, hardcoded into the rendering pipeline. However, the colormap handling in future will belong in the GUI layer, allowing for real-time interaction and customization. This will be achieved by passing colormap handles or indices through the Shader Binding Table (SBT), enabling dynamic transfer function control directly from the user interface without recompilation. In figure  \ref{fig:collage_part2}, two volumetric datasets from OpenVDB have been used for showing application of SciVis style transfer functions.

\subsubsection{Combining Volumes and Level-sets}
\begin{figure}[tbp]
  \centering
  \includegraphics[width=0.95\columnwidth, alt={Big letter C on a gray background.}]{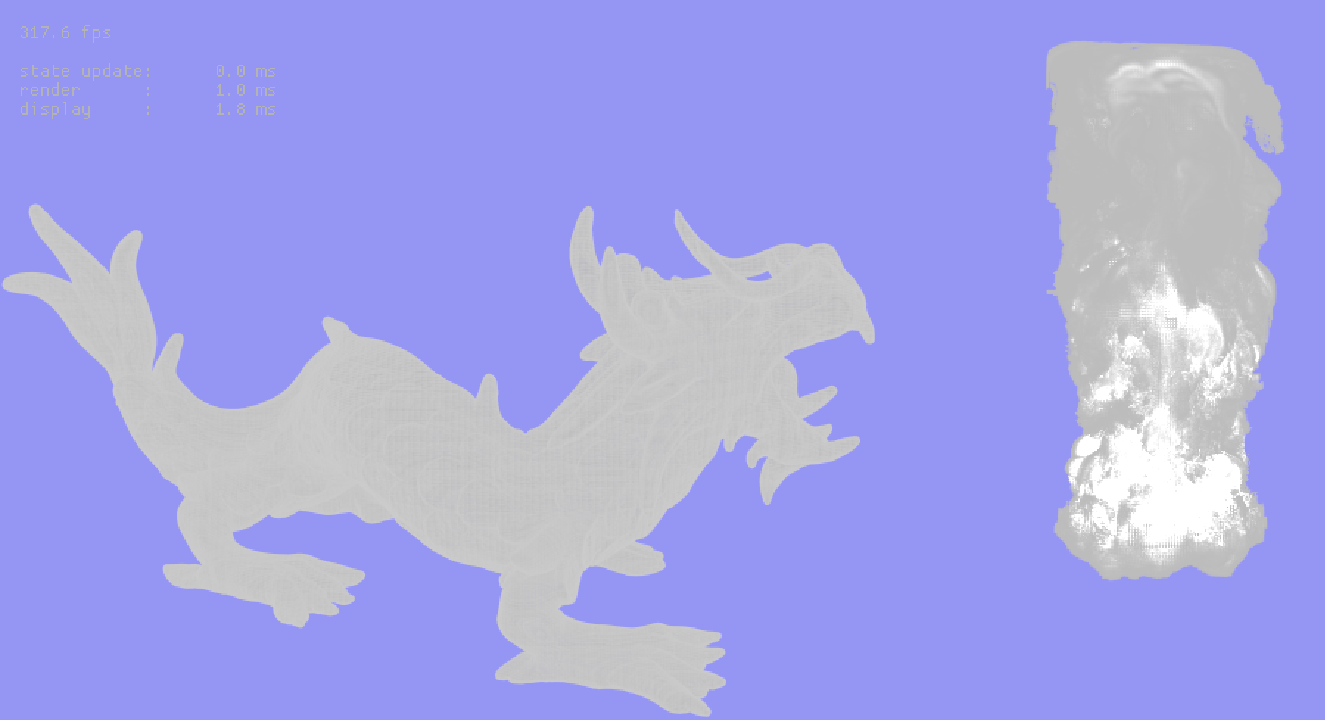}
  \caption{Combined surface and volume modeling with 3D Gaussians}
  \label{fig:comb-surf-vol}
\end{figure}

Gaussians can represent both surfaces and volumes alike. Unlike traditional rendering pipelines which treat volumes as scalar fields and surfaces as polygonal meshes, often requiring distinct data structures, Gaussian-based approach allows both types of structures to be encoded similarly. We still require separate shaders when doing photorealistic rendering, but for SciVis use cases having a unified hybrid renderer could support some datasets. See figure \ref{fig:comb-surf-vol}.

\section{Results}
We evaluate our initial Gaussian-based renderer using six publicly available datasets from the OpenVDB repository: bunny\_cloud, explosion, fire, utahteapot, venusstatue and dragon\cite{openvdb_download}. The first three datasets are volumetric and remaining are level-sets. Each dataset was processed into 3D Gaussian primitives across three LOD to test performance and visual quality trade-offs.

The three LOD modes are defined as follows:

\begin{itemize}
  \item \textbf{Low LOD}: Each leaf node in the OpenVDB grid is represented by a single Gaussian, whose shape (sphere or ellipsoid) reflects the distribution of active voxels within the node. This setting produces the smallest number of Gaussians and yields fast rendering, but the output lacks fine detail and tends to appear overly crude.

  \item \textbf{Medium LOD}: Dense and sparse leaf nodes are processed by merging all active voxels into the largest possible 4x4x4 groups and 2×2×2 groups respectively, while the remaining active voxels are added as individual voxel-sized Gaussians. This setting provides a balance between performance and detail, capturing medium-scale structures well.

  \item \textbf{High LOD}: All sparse leaf nodes contribute one Gaussian per active voxel, while dense leaf nodes are merged into 2×2×2 blocks when possible. This configuration preserves fine structural details at the cost of a much higher Gaussian count and memory usage.
\end{itemize}

At all LODs, the rendering exhibits a distinctive partially-segmented appearance, which is a characteristic of using non-overlapping Gaussians. Each Gaussian occupies a distinct spatial region and without proper blending it can lead to visible seams and particle boundaries. While this design simplifies data management and GPU traversal, it introduces a certain degree of visual discontinuity. 

Our current implementation uses a closest-hit shader, which only processes the first intersection between a ray and a Gaussian. To simulate volumetric integration, we manually advance the ray beyond the first intersection using recursive \texttt{optixTrace} calls. However, this approach has resulted in sub-optimal accumulation of transparency and color. This is likely due to limitations in carrying data across recursive shader calls and inconsistent resetting of payload values(OptiX 32-bit registers) between manual \texttt{optixTrace} calls. Achieving correct volume compositing requires integrating opacity contributions from all Gaussians along the ray path. This is difficult to be achieved with a closest-hit-only approach in our experience.

To address this, we plan to adopt an any-hit shader mechanism in future iterations. This would allow the renderer to detect all intersected Gaussians per ray and accumulate transmittance and opacity correctly, essential for SciVis effects like  smooth blending and volumetric attenuation.

In the medium and high LOD settings, a significant number of single-voxel Gaussians are generated, as a spatial clustering algorithm for grouping leftover voxels is missing. After attempting to chunk voxels into 2×2×2 or 4×4×4 blocks within a sparse leaf node, some voxels inevitably remain ungrouped. These leftover voxels are often spatially scattered, making it inappropriate to simply average their positions and create a single ellipsoidal Gaussian as done in the low LOD case. Without considering their spatial distribution, such a centroid-based approach would lead to inaccuracies in representation. Proximity and orientation of voxels in 3D space should enable the clustering. Such a spatially-aware clustering algorithm is non-trivial and will be the focus of future iterations.

 OptiX’s hardware-accelerated bounding box (AABB) intersections and ray marching provides good performance to our renderer. The renderer consistently provides above 150 FPS of rendering performance for all datasets tested in all LOD. Even in higher LOD settings, where the scene contains millions of Gaussians, the renderer maintains real-time performance with the tested datasets. These results confirm that representing volumes using optimized Gaussian primitives and bounding volumes can offer significant speedups over traditional dense-grid volume rendering, especially when combined with hardware-accelerated ray tracing frameworks.

Regarding the memory footprint of the Gaussian points, we account for the storage requirements of each point's global position, covariance values and opacity across different LOD and  different type of datasets:

\begin{itemize}
  \item Global Position: Each Gaussian's position is stored as a 3D coordinate.
  \item Covariance Values: The representation varies based on the Gaussian's shape:
  \begin{itemize}
    \item Spheres: Require a single floating-point value.
    \item Ellipsoids: Require three floating-point values.
  \end{itemize}
   \item Opacity Values: Volumes have a scalar field of opacity or density values, where the opacity varies throughout the volume. On the other hand, level sets are represented as signed-distance fields, where the opacity is uniform across the surface. Therefore, for level sets, such as the utahteapot or venusstatue, we can avoid storing a separate opacity value for each Gaussian.
\end{itemize}

The ratio of spheres to ellipsoids influences the overall memory consumption. Additionally, since each point is assigned its own AABB in OptiX, we do not need to explicitly store information about point counts or offsets, further optimizing memory usage.

Our rendering output for different VDB datasets is presented on the last page(see figure \ref{fig:big_collage}). 
In the table of results(\ref{tab:gaussian_datasets}), the binary files from OpenVDB repositories have their real file size mentioned in the first column, followed by voxel count in millions(M). Then we have the three detail settings, each with their memory footprint in MegaBytes(MB) and number of Gaussians(in thousands, K; or in millions, M).

\begin{table}[tb]
  \caption{Summary of Gaussian-converted volumetric datasets. Each LOD column shows file size (MB) and number of Gaussians (K).}
  \label{tab:gaussian_datasets}
  \scriptsize
  \centering
  \setlength{\tabcolsep}{3.5pt}
  \definecolor{lightgray}{rgb}{0.9, 0.9, 0.9}
  \begin{tabular*}{\columnwidth}{@{\extracolsep{\fill}} lcc
    >{\centering\arraybackslash}m{1.15cm}
    >{\centering\arraybackslash}m{1.15cm}
    >{\centering\arraybackslash}m{1.15cm}}
    \toprule
    Dataset & Size (MB) & Voxels &
    \makecell{Low LOD\\(MB \\ \#G )} &
    \makecell{Mid LOD\\(MB \\ \#G )} &
    \makecell{High LOD\\(MB \\ \#G )} \\
    \midrule
    bunny-cloud   & 74  & 144M   & \makecell{2.8MB\\66K} & \makecell{97MB\\3.5M} & \makecell{778MB\\12M} \\
    \rowcolor{lightgray}
    explosion     & 72  & 12M    & \makecell{0.4MB\\10.2K} & \makecell{9.8MB\\366K} & \makecell{42.9MB\\1.5M} \\
    fire          & 23  & 8M  & \makecell{0.68MB\\12.5K} & \makecell{24.5MB\\898K} & \makecell{88.4MB\\3.2M} \\
    utahteapot    & 12  & 280M   & \makecell{1.6MB\\35K} & \makecell{51.2MB\\2.1M} & \makecell{163MB\\6.9M} \\
    venusstatue   & 20  & 130M   & \makecell{1.4MB\\29K} & \makecell{40.8MB\\1.7M} & \makecell{135MB\\5.7M} \\
    dragon        & 45  & 2.4B & \makecell{5.8MB\\124K} & \makecell{174.4MB\\7.4M} & \makecell{546MB\\23.3M} \\
    \midrule
    \bottomrule
  \end{tabular*}
\end{table}

\begin{figure*}[tbp]
  \centering
  \setlength{\tabcolsep}{2pt} % space between columns
  \renewcommand{\arraystretch}{0.1} % vertical spacing

  \begin{tabular}{ccc}
      \includegraphics[width=0.31\textwidth]{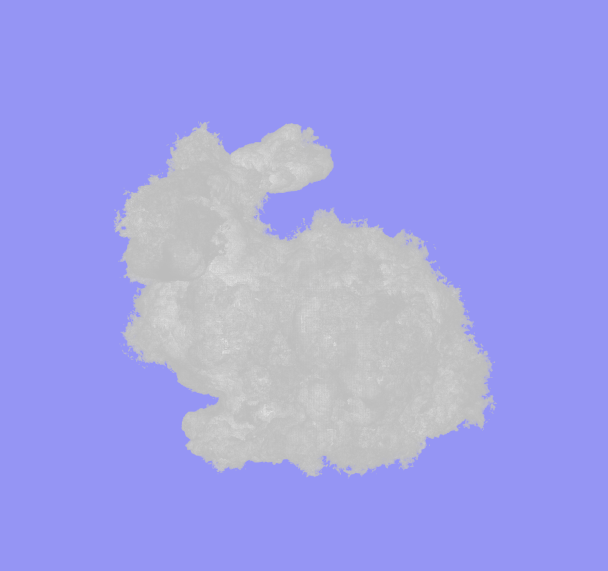} &
    \includegraphics[width=0.31\textwidth]{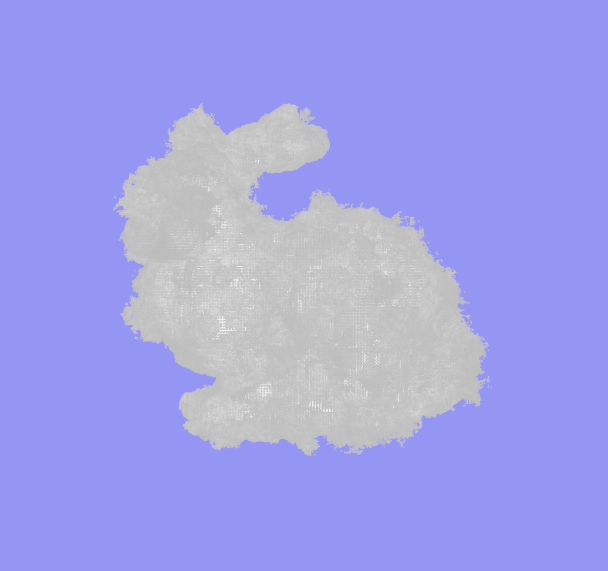} &
    \includegraphics[width=0.31\textwidth]{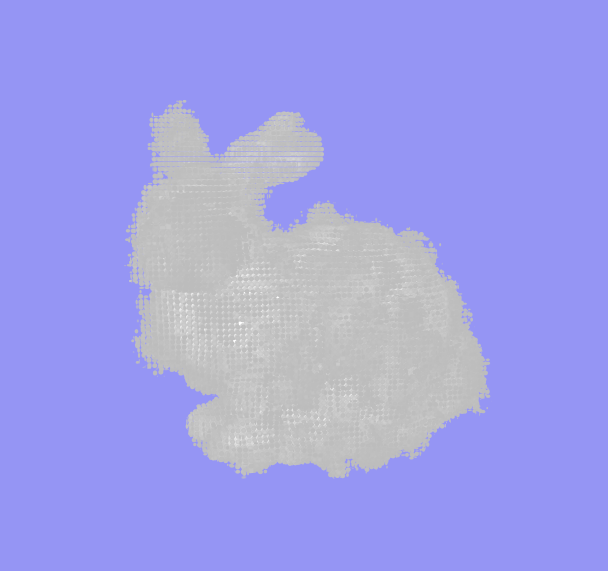} \\
    
    \includegraphics[width=0.31\textwidth]{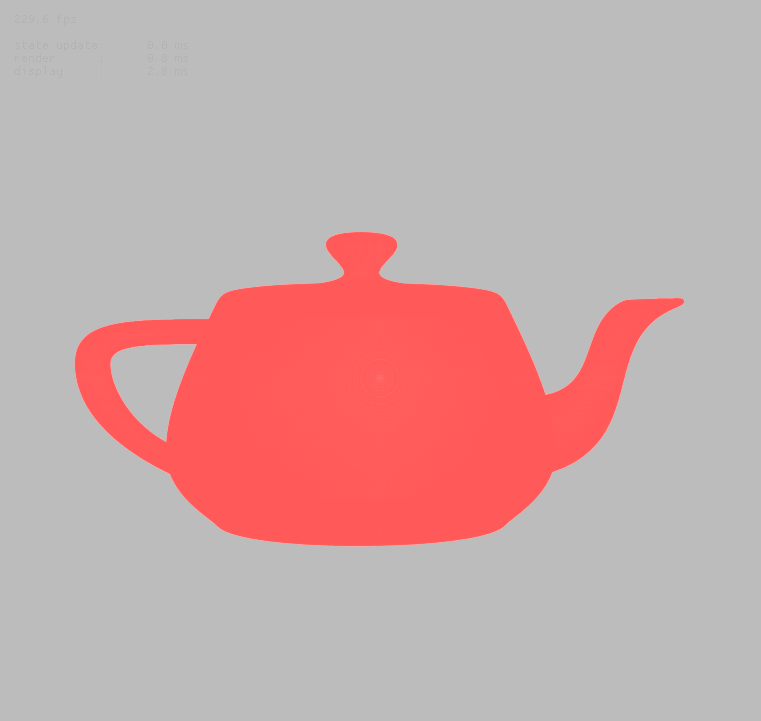} &
    \includegraphics[width=0.31\textwidth]{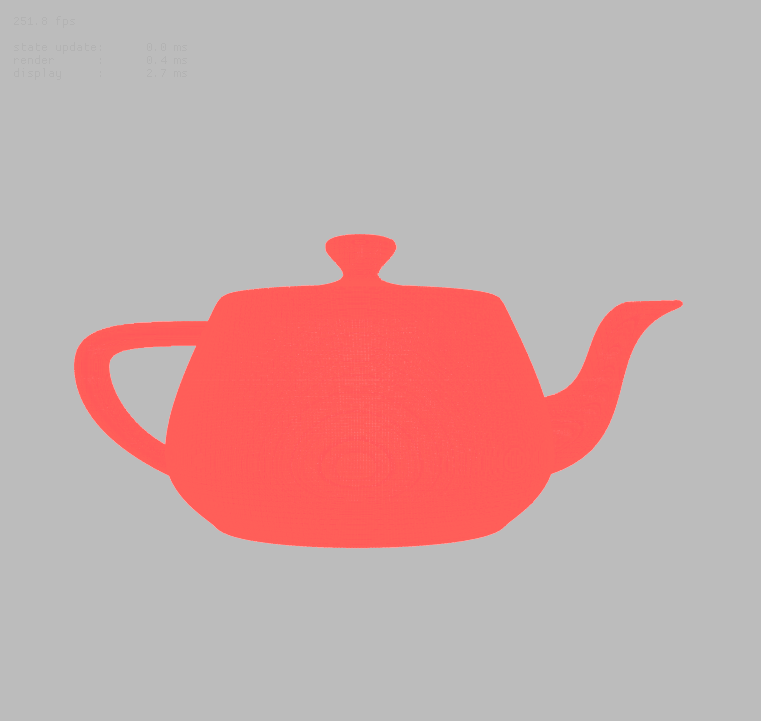} &
    \includegraphics[width=0.31\textwidth]{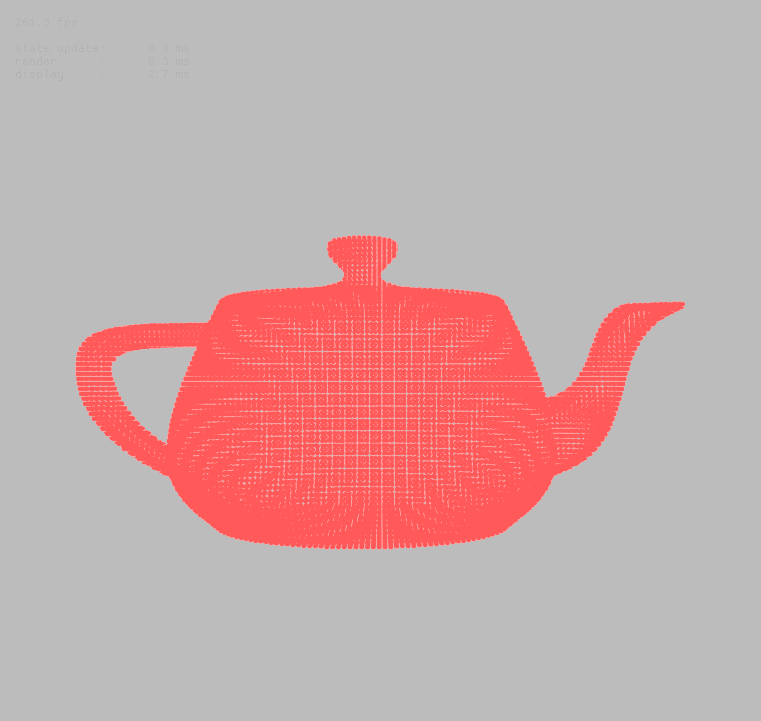} \\

    \includegraphics[width=0.31\textwidth]{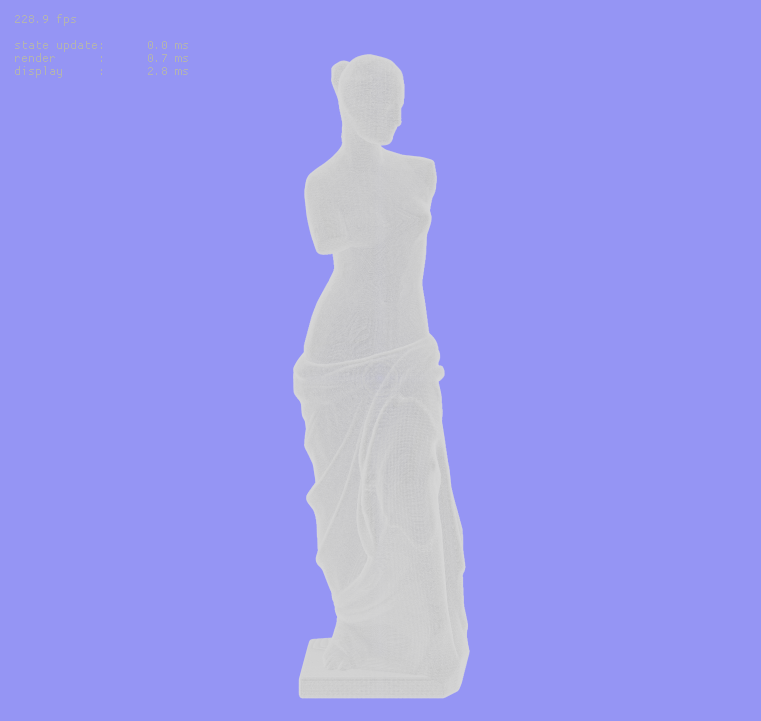} &
    \includegraphics[width=0.31\textwidth]{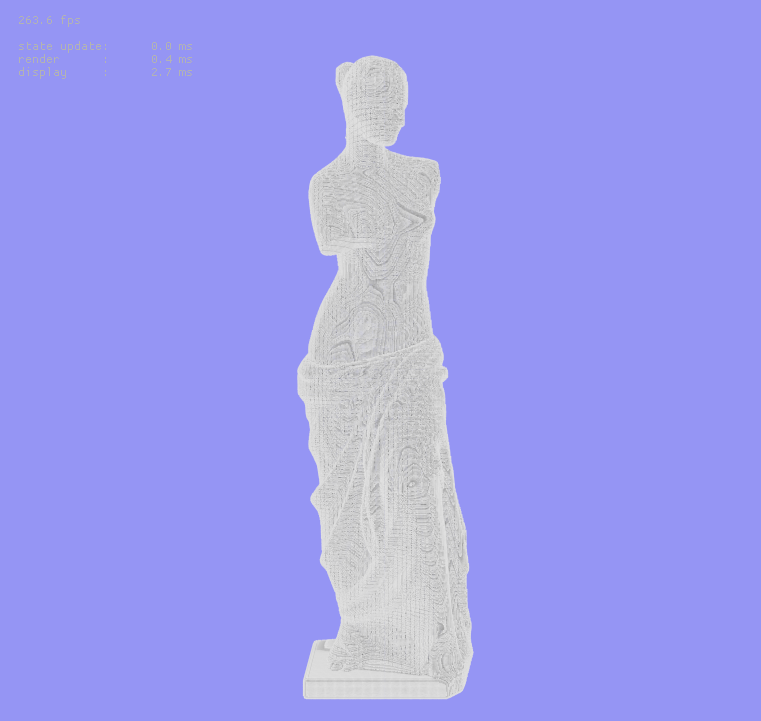} &
    \includegraphics[width=0.31\textwidth]{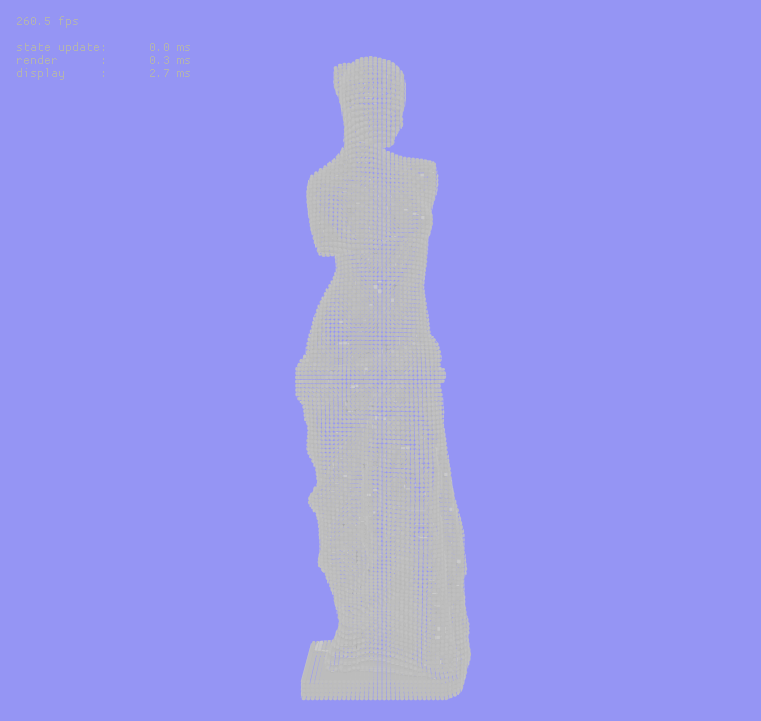} \\

    \includegraphics[width=0.31\textwidth]{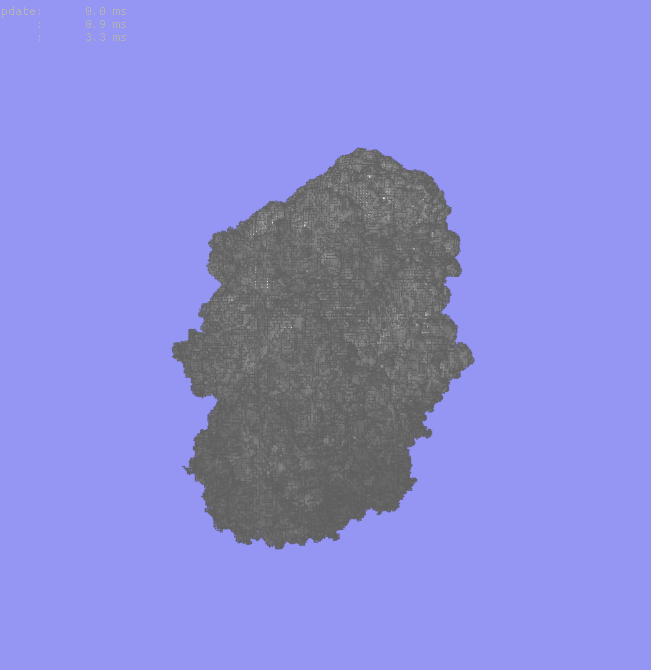} &
    \includegraphics[width=0.31\textwidth]{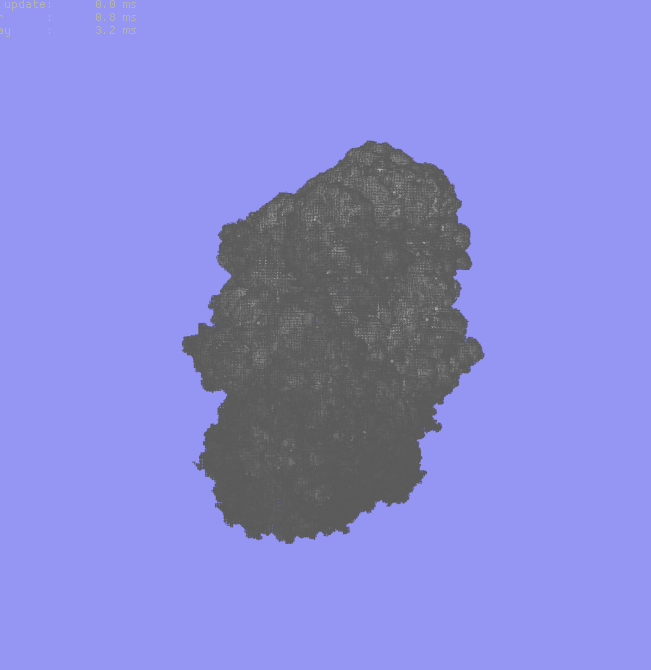} &
    \includegraphics[width=0.31\textwidth]{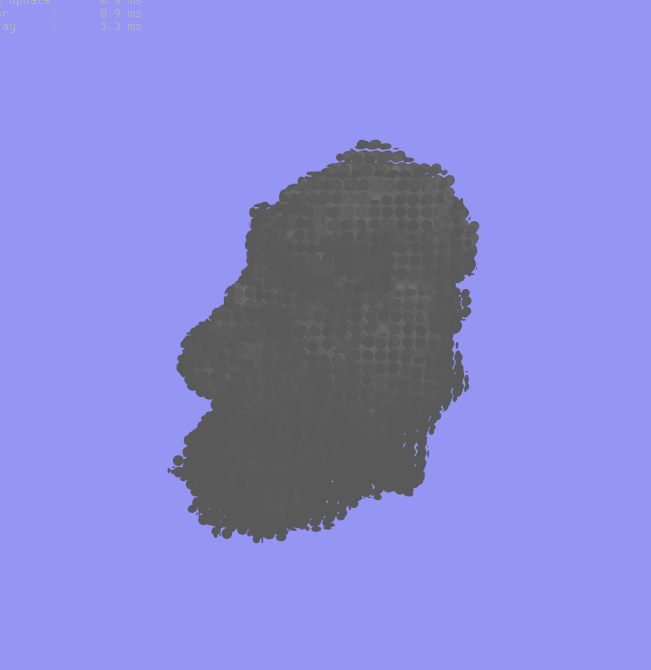} \\

  \end{tabular}

  \caption{Top to Bottom: bunny\_cloud, utahteapot, venusstatue and explosion in three LOD setting i.e., left-to-right: High, Medium and Low}
  \label{fig:big_collage}
\end{figure*}
\section{Conclusion}
In the result we highlighted, the 'explosion' dataset performs very well to our reduced representation scheme across different LOD and it is attributed to the \emph{higher ratio of dense to non-dense nodes}. We would like to conclude this paper by noting that finding a one-size-fits-all Gaussian generation scheme for various VDB datasets will require further careful iterations and the testing of more advanced clustering algorithms, which themselves have a rich history of research.

Additionally, adopting the VDB volumetric format for data storage could significantly benefit SciVis applications, as it offers a memory-efficient method for storing density and other scalar fields. Medical datasets, such as MRI or CT scans, often contain large portions of irrelevant data corresponding to the surrounding area of the scanned object. Efficient writers could convert these datasets into VDB files, drastically reducing memory usage. Our method demonstrates that further compression is possible depending on the required LOD. Such a combination is especially suitable for handling extremely large datasets. 

Lastly, the power of modern GPUs has made it possible to efficiently trace millions of particles, making voxel-based volume representations appear rigid in comparison, particularly for systems that require greater flexibility in representation, such as unstructured grids with polygonal cells.

\section{Future Work}
While our current renderer demonstrates the potential of representing volumetric data using Gaussian primitives, several areas remain open for future improvement:
\begin{itemize}
  \item
    Overlapping Gaussian Representation: Currently, all Gaussians are treated as non-overlapping entities, resulting in visual artifacts such as blockiness and discontinuities. Supporting overlapping Gaussians would enable smoother visual transitions, better density accumulation and more faithful reconstruction of soft volume boundaries. This will require changes to how Gaussians are stored, blended and intersected during rendering.
    \item
    Intelligent and Spatially Aware Clustering: Our current LOD construction is limited to simple voxel chunking (e.g., 2×2×2 or 4×4×4) without considering the spatial distribution of ungrouped voxels. In future work, we plan to implement a spatially aware clustering algorithm that can intelligently group remaining voxels based on proximity and structure, thereby reducing Gaussian count while preserving accuracy. Techniques such as voxel adjacency graphs\cite{Nourian2023VoxelGraphOperators} or density-based clustering (e.g., DBSCAN\cite{Ester1996DensityBased}) are promising directions.
\item
    Improved Transparency and Color Accumulation: Accurate volumetric rendering requires integrating contributions from multiple overlapping elements along a ray. At present, our renderer relies on a closest-hit shader and manual recursive ray traversal, which is somewhat limited by OptiX's 8-payload registers and inconsistent behavior during manual trace progressions. Future iterations will explore any-hit shaders to ensure better transmittance and color accumulation. These changes are necessary to achieve correct volumetric effects such as soft fading and smooth opacity blending.
  \end{itemize}

%% if specified like this the section will be omitted in review mode
\acknowledgments{%
	The authors wish to thank Martin Reiss of AKAD University Stuttgart for proofreading the paper and Will Usher of Luminary Cloud for discussing the OptiX ray tracing pipeline features. 
}

\bibliographystyle{abbrv-doi-hyperref}

\bibliography{vis}

\newpage
\appendix 
\section*{Appendix}

\subsection*{Ray-Gaussian intersection}
The intersection of the ray with the Gaussian can be computed as follows:

\begin{align*}
\text{Mahalanobis Distance: } & \quad (\mathbf{x} - \mu)^T \Sigma^{-1} (\mathbf{x} - \mu) \\
\text{Substitute } \mathbf{x} = \mathbf{o} + \tau \mathbf{d}: & \quad (\mathbf{o} + \tau \mathbf{d} - \mu)^T \Sigma^{-1} (\mathbf{o} + \tau \mathbf{d} - \mu) \\
& = (\mathbf{o} - \mu)^T \Sigma^{-1} (\mathbf{o} - \mu) + 2\tau (\mathbf{o} - \mu)^T \Sigma^{-1} \mathbf{d} + \tau^2 \mathbf{d}^T \Sigma^{-1} \mathbf{d} \\
\text{Set to unit sphere: } & \quad (\mathbf{o} - \mu)^T \Sigma^{-1} (\mathbf{o} - \mu) + 2\tau (\mathbf{o} - \mu)^T \Sigma^{-1} \mathbf{d} + \tau^2 \mathbf{d}^T \Sigma^{-1} \mathbf{d} = 1 \\
\text{Rearrange to quadratic form: } & \quad A \tau^2 + B \tau + C = 1 \\
\text{where: } & \quad A = \mathbf{d}^T \Sigma^{-1} \mathbf{d}, \quad B = 2 (\mathbf{o} - \mu)^T \Sigma^{-1} \mathbf{d}, \quad C = (\mathbf{o} - \mu)^T \Sigma^{-1} (\mathbf{o} - \mu) - 1 \\
\text{Solve for } \tau: & \quad \tau_{1,2} = \frac{-B \pm \sqrt{B^2 - 4AC}}{2A}
\end{align*}
\end{document}